# Towards a Pattern Language for Quantum Algorithms

Frank Leymann[0000-0002-9123-259X]

IAAS, University of Stuttgart,

Universitätsstr. 38, 70569 Stuttgart, Germany
`Frank.Leymann@iaas.uni-stuttgart.de`

**Abstract.** Creating quantum algorithms is a difficult task, especially for computer scientist not used to quantum computing. But quantum algorithms often use similar elements. Thus, these elements provide proven solutions to recurring problems, i.e. a pattern language. Sketching such a language is a step towards establishing a software engineering discipline of quantum algorithms.

## 1. Introduction

### 1.1. Patterns and Pattern Languages

There is a significant difference in how quantum algorithms are presented and invented, and the way how traditional algorithms are build. Thus, computer scientists and software developers used to solve classical problems need a lot of assistance when being assigned to build quantum algorithms.

To support and guide people in creating solutions in various domains, pattern languages are established. A *pattern* is a structured document containing an abstract description of a proven solution of a recurring problem. Furthermore, a pattern points to other patterns that may jointly contribute to an encompassing solution of a complex problem. This way, a network of related patterns, i.e. a *pattern language*, results.

This notion of pattern and pattern language has its origin in [1]. Although invented to support architects in building houses and planning cities, it has been accepted in several other domains like pedagogy, manufacturing, and especially in software architecture (e.g. [13]).

In this paper, we lay the foundation for a pattern language for quantum algorithms. The need for documenting solutions for recurring problems in this domain can be observed in text books like [19, 21] that contain unsystematic explanations of basic „tricks" used in quantum algorithms. Our contribution is to systematize this to become a subject of a software engineering discipline for quantum algorithms.

### 1.2. Structure of a Pattern Document

A document specifying a pattern within a certain domain follows a fixed structure. While this structure may vary from domain to domain, many elements are in common, i.e. independent of the domain:

Each pattern has a *name*. This name should be descriptive, identifying the problem to be solved.

The *intend* of the pattern briefly describes the goal to be achieved with the solution described by the pattern.

An *icon* represents the pattern visually, e.g. as a mnemonic. While the gate model makes heavy use of icons (as quantum gates and their wiring), other models don't; pattern icons may add a visual aspect to such models. Even for the gate model, patterns often abstract gates and their compositions, i.e. pattern icons are a more abstract representation of parts of an algorithm.

The *problem statement* concisely summarizes the problem solved by the pattern. Based on this, the reader can immediately decide whether the pattern is relevant for the problem at hand to be solved.

The *context* describes the situation or the forces, respectively, that led to the problem. It may refer to other patterns already applied.

The *solution* is the most important element of the pattern: it specifies in an abstract manner how to solve the problem summarized in the problem statement. The problem statement together with the solution is the underpinning of the pattern. Variants of solutions may be described depending on different flavors of the context.

The *know uses* section refers to algorithms that make use of the pattern. It confirms that the problem is recurring and that the presented solution is proven.

Other patterns related to the current one are referenced in the *next* element. These references link the individual patterns into a pattern language.

### 1.3. Overview

The patterns we propose in section 2 are derived from algorithms based on the gate model. This is not a restriction in principle, because patterns based on other models may be added in future. Furthermore, patterns based on one model may be transformed into equivalent patterns of other model (e.g. the gate model is know to be equivalent to the measurement-based model [16]). Finally, we briefly indicate in section 3 how patterns may be used in developing quantum algorithms.

## 2. Patterns for Quantum Algorithms

In this section we describe an initial set of basic patterns and their relations. Note, that this pattern language is far from being encompassing, i.e. it is expected that this pattern language evolves over time.

## 2.1. Initialization (aka State Preparation)

**Intend**: At the beginning of a quantum algorithm, the quantum register manipulated by the algorithm must be initialized. The initialization must be as easy as possible, considering requirements of the steps of the algorithm.

> 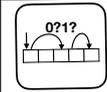 *How can the input of a quantum register be initialized in a straight-forward manner, considering immediate requirements of the following steps of the quantum algorithm?*

**Context**: An algorithm typically requires input representing the parameters of the problem to be solved. Most quantum algorithms encode this input as part of the unitary transformations making up the quantum algorithm. E.g. if the overall algorithm is $U=U_n \circ \ldots \circ U_i \circ U_{i-1} \circ \ldots \circ U_1$, then $U_1,\ldots,U_{i-1}$ are operators that furnish the register to hold the parameters of the problem solved by the following operators $U_i, \ldots, U_n$. However, the initial state operated on by $U_{i-1} \circ \ldots \circ U_1$ must be set; $U_{i-1} \circ \ldots \circ U_1$ is called *state preparation* [25].

**Solution**: Often, the register will be initialized as the unit vector $|0\ldots0\rangle$. This register may have certain ancilla bits or workspace bits distinguished that are used to store intermediate results, to control the processing of the algorithm etc..

For example, the register is initialized with $|0\rangle^{\otimes n} |0\rangle^{\otimes m}$ (where the second part of the register consists of workspace bits) in order to compute the function table of the Boolean function $f : \{0,1\}^n \to \{0,1\}^m$.

An initialization with $|0\rangle^{\otimes n} |1\rangle$ supports to reveal membership in a set which is defined based on an indicator function (used to solve decision problems, for example) by changing the sign of the qbits representing members of this set.

Based on these simple initializations, more advanced states can be prepared. For example, [7] discusses several algorithms to load classical bits into a quantum register. [25] presents how to load a complex vector, [8] how to load a real vector based on corresponding data structures; thus, a matrix can be loaded as a set of vectors [18].

**Known uses**: All algorithms must be initialized somehow.

**Next**: Often, after initialization the register must be brought into a state of *uniform superposition*. *Function tables* require the initializations discussed here. An initialized register may become input to an *oracle*.

## 2.2. Uniform Superposition

**Intend:** Typically, the individual qbits of a quantum register have to be in multiple states at the same time without preferring any at these states at the beginning of the computation.

*How can an equally weighted superposition of all possible states of the qbits of a quantum register be created?*

**Context**: One origin of the power of quantum algorithms stems from quantum parallelism, i.e. the ability of a quantum register to represent multiple values at the same time. This is achieved by bringing (a subset of) the qbits of a quantum register into superposition. Many algorithms assume that at the beginning this superposition is uniform, i.e. the probability of measuring any of the qbits is the same.

**Solution**: Uniform superposition is achieved by initializing the quantum register as the unit vector $|0...0\rangle$ and applying the Hadamard transformation afterwards:

$$H^{\otimes n}\left(|0\rangle^{\otimes n}\right) = \frac{1}{\sqrt{2^n}} \sum_{x=0}^{2^n-1} |x\rangle$$

In case the quantum register includes ancilla bits or workspace bits in addition to the computational basis, the computational basis is brought into superposition as described. The other bits may be brought into superposition themselves or not. This is achieved by using a tensor product $H^{\otimes n} \otimes U$, where $H^{\otimes n}$ operates on the computational basis and U operates on the other bits (e.g., U=I in case the other bits are not brought into superposition).

**Known uses**: Most algorithms make use of uniform superposition.

**Next**: Creating uniform superposition makes use of *initialization*. A register in uniform superposition may be *entangled*. A register in uniform superposition may be input to an *oracle*.

### 2.3. Creating Entanglement

**Intend**: A strong correlation between qbits of a quantum register is often needed in order to enable algorithms that offer a speedup compared to classical algorithms.

*How can an entangled state be created?*

**Context**: Entanglement is one of the causes of the power of quantum algorithms (see [5], although entanglement is not a necessity [3]). A quantum algorithm showing exponential speedup requires entanglement [17]. Thus, after initialization of a quantum register it should often be entangled for its further processing.

**Solution**: Several approaches can be taken to create an entangled state. For example, assume a binary function f:$\{0,1\}^n \to \{0,1\}^m$ and the corresponding unitary operation

$$U_f : \{0,1\}^{n+m} \to \{0,1\}^{n+m}, \quad U_f(|x,y\rangle) = |x, y \oplus f(x)\rangle.$$

Then, the following state is entangled.

$$U_f\left(H^{\otimes n} \otimes I^{\otimes m}\right)\left(|0\rangle \otimes |0\rangle\right).$$

With f=id it is $U_{id}$ = CNOT, which shows that CNOT((H⊗I)(|0>⊗|0>)) is entangled.

**Known uses**: Many algorithms make use of entanglement.

**Next**: Typically, *initialization* precedes the creation of entanglement. A f*unction table* results from the above creation of entanglement based on $U_f$.

### 2.4. Function Table

**Intend**: Some problems can be reduced to determining global properties of a function. For that purpose, the corresponding function table should be computed efficiently and made available for further analysis.

| x | f(x) |
|---|---|

*How can a function table of a finite Boolean function be computed?*

**Context**: In order to compute the function table of a function

$$f : \{0,1\}^n \rightarrow \{0,1\}^m,$$

a classical algorithm requires to invoke the function for each value of the domain. Quantum parallelism allows to compute the values of such a finite Boolean function as a whole in a single step. This can be used to speedup finding global properties of the corresponding function. Note, that in case m=1 the Boolean function is often an indicator function used to determine solutions of a decision problem.

**Solution**: The quantum register is split into the computational basis (the domain of the function f) consisting of n qbits x, and a workspace consisting of m qbits y, which is used to hold the values of f. Based on this, the unitary operator

$$U_f\ |x,y\rangle = |x,\ y \oplus f(x)\rangle$$

is defined.

After initializing the register with $|0\rangle^{\otimes n}|0\rangle^{\otimes m}$, the computational basis is brought into uniform superposition via $H^{\otimes n}$ leaving the workspace unchanged, and then the operator $U_f$ is applied only once resulting in the function table:

$$|0\rangle^{\otimes n}|0\rangle^{\otimes m} \xmapsto{H^{\otimes n}\otimes I} \left(\frac{1}{\sqrt{2^n}}\sum_x |x\rangle\right)\otimes |0\rangle^{\otimes m} \xmapsto{U_f} \frac{1}{\sqrt{2^n}}\sum_x |x\rangle|f(x)\rangle$$

In case of an indicator function f (e.g. if f is representing a decision problem), the register is initialized with $|0\rangle^{\otimes n}|1\rangle$. Uniform superposition of the complete register is furnished by $H^{\otimes n+1}$. Applying $U_f$ finally results in

$$|0\rangle^{\otimes n}|1\rangle \overset{H^{\otimes n}\otimes H}{\mapsto} \left(\frac{1}{\sqrt{2^n}}\sum_x |x\rangle\right)\otimes |-\rangle \overset{U_f}{\mapsto} = \left(\frac{1}{\sqrt{2^n}}\sum_{x=0}^{2^n-1}(-1)^{f(x)}|x\rangle\right)\otimes |-\rangle$$

Thus, members of the computational basis indicate by their sign whether they are detected by the indicator function (minus sign) or not (plus sign) - aka „phase kickback".

**Known uses**: The algorithms of Deutsch, Deutsch-Jozsa, Grover, Shor and others make use of function tables.

**Next**: Function tables require *initialization* discussed before. *Uniform superposition* of the computational basis is established before the function table is computed. *Amplitude amplification* is a generalization of function tables. The computation is performed by an *oracle*. Often, *uncompute* is required to continue processing.

### 2.5. Oracle (aka Black Box)

**Intend**: Quantum algorithms often need to compute values of a function f without having to know the details how such values are computed.

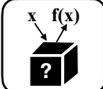
*How can the computation of another quantum algorithm be reused?*

**Context**: Divide-and-Conquer is a well-established method in computer science to simplify the solution of complex problems. The concept of an oracle (or black box) as a granule of reuse with hidden internals supports this method for building quantum algorithms.

**Solution**: Oracles are used in problem specific manners. [14] discusses various kinds of oracles. Limitations of using oracles are discussed in [26].

**Known uses**: The algorithms of Deutsch, Deutsch-Jozsa, Bernstein-Vazirani, Simon, Grover and others make use of oracles. See [20] for further usages.

**Next**: An oracle often requires to *uncompute* its result state, and assumes a properly prepared register as input (*initialization*).

### 2.6. Uncompute (aka Unentangling aka Copy-Uncompute)

**Intend**: Often, entanglement of the computational basis of a quantum register with temporary qbits (ancilla, workspace) has to be removed to allow proper continuation of an algorithm.

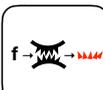
*How can entanglement be removed that resulted from a computation?*

**Context**: A computation often needs temporary qbits, and at the end of the computation these qbits are entangled with the computational basis. This hinders access to the actual result of the computation, especially if the computation was used just as an intermediate step within an algorithm.

For example, if the computation should produce $\sum \alpha_i |\varphi_i\rangle$ but in fact it produces $\sum \alpha_i |\varphi_i\rangle |\psi_i\rangle$, the temporary qbits $|\psi_i\rangle$ can not be simply eliminated unless

$$\sum \alpha_i |\varphi_i\rangle |\psi_i\rangle = \left(\sum \alpha_i |\varphi_i\rangle\right) \otimes |\psi_i\rangle,$$

i.e. unless the computational basis and the temporary qbits are separable.

**Solution**: Most algorithms map |x>|0> |0> to |x>|g(x)>|f(x)> to compute a function f [8]. I.e. the second qbits represent a workspace that contains garbage g(x) at the end of the computation. This garbage has to be set to |0> to allow for proper continuation, especially if future parts of the algorithm expects the workspace to be initialized again by |0>.

More precisely, assume the computation $U_f$ resulted in

$$|x\rangle|0\rangle|0\rangle \xmapsto{U_f} \sum_y a_y |x\rangle|y\rangle|f(x)\rangle,$$

i.e. the garbage state is $|g(x)\rangle = \sum_y a_y |y\rangle$. Now, a fourth register initialized to |0> is added, and CNOT is applied (bitwise) to this fourth register controlled by the third register: this copies f(x) to the fourth register and $\sum_y a_y |x\rangle|y\rangle|f(x)\rangle|f(x)\rangle$ results.

Next, $U_f^{-1}$ is applied to the first three registers, giving |x>|0>|0>|f(x)>. Then, SWAP is applied to the last two registers leaving |x>|0>|f(x)>|0>. This now allows to discard the last register leaving |x>|0>|f(x)> as wanted (more details in [8]). [23] discusses how to use uncompute in several situations.

**Known uses**: Deutsch-Joza, the HHL algorithm [15], quantum walks, realizations of classical circuits as quantum algorithms etc make use of uncompute.

**Next**: An *oracle* often produces a state that is an entanglement between the computational basis and some temporary qbits, thus requires uncompute. A *function table* may be seen as a special case of an oracle.

### 2.7. Phase Shift

**Intend**: In a given register certain qbits should be emphasized.

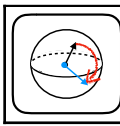 *How can important aspects of a state been efficiently distinguished?*

**Context**: When an algorithm is applied iteratively, and each iteration is assumed to improve the solution, those parts of the solution that did improve should be indicated. A phase shift can be such an indication.

**Solution**: The following operator $S_G^\varphi$ can be efficiently implemented (see [24]) in terms of number of gates used:

$$\sum_{x=0}^{N-1} a_x |x\rangle \overset{S_G^\varphi}{\mapsto} \sum_{x \in G} e^{i\varphi} a_x |x\rangle + \sum_{x \notin G} a_x |x\rangle$$

This operator shifts the qbits in G ⊆ {0,…,N-1} (the qbits improved: „good set") by phase $\varphi$ and leaves the other qbits unchanged. There is even a variant of the operator that shifts the phases of the qbits in the good set by different values, i.e. $\varphi=\varphi(x)$.

**Known uses**: The algorithms of Grover, Deutsch-Jozsa etc. use a phase shift.

**Next**: A *function table* based on an indicator function is a phase shift, with G as the set of base vectors qualifying under the indicator function. An *amplitude amplification* makes use of two phase shifts. A phase shift is used as an *oracle*.

## 2.8. Amplitude Amplification

**Intend**: Based on an approximate solution, the probability to find the precise solution should be increased from run to run of an algorithm U.

*How can the probability of finding a solution be inreased?*

**Context**: The function table of an indicator function f may list all solutions of a problem (i.e. f(x) = 1 ⇔ x solves the problem). By measuring the corresponding state, a solution is found with a certain probability. But measuring destroys the state, i.e. if a solution is not received by measurement, the computation has to be performed again to support another new measurement.

Thus, a mechanism is wanted that doesn't need measurements and that allows to continue with the state achieved in case a solution is not found.

**Solution**: State is transformed in such a way that values of interest get a modified amplitude such that they get a higher probability of being measured after a couple of iterations [4].

The phase shift $S_G^\pi$ changes the sign of the phase of elements in G, the phase shift $S_0^\pi$ changes the sign of $|0\rangle$ (the start value of the iteration) and leaves the other elements unchanged. Let U be the algorithm for computing approximate solutions (not using any measurements). Define the following is unitary operation:

$$Q = -U S_0^\pi U^{-1} S_G^\pi$$

If U is an algorithm that succeeds with a solution with probability t, 1/t iterations are required on the average to find a solution. U|0> is assumed to have a non-zero

amplitude in G, otherwise no speedup can be achieved. If U has this property, Q will produce a solution within $O\left(\sqrt{1/t}\right)$ iterations - which is a quadratic speedup. The number of iterations to be performed with Q is about

$$\frac{\pi}{4} \cdot \frac{1}{\left|P_G U|0\rangle\right|}$$

where $P_G$ is the projection onto the subspace spanned by G.

**Known uses**: The algorithms of Grover and Simons, for example, make use of amplitude amplification. Also, the HHL algorithm for solving linear equations [15] uses this pattern. The state preparation algorithm of [25] uses amplitude amplification too. [4] discusses more algorithms making use of it.

**Next**: Part of the unitary operation Q is the *function table* $S_G^\pi$, which is also a special case of a *phase shift*. Amplitude amplifications are used as *oracle*s.

## 2.9. Speedup via Verifying

**Intend**: Verifying whether a claimed solution is correct or not is sometimes simple. Such verifications may then be used to speedup solving a corresponding problem.

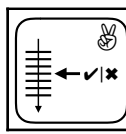 *How can a speedup be achieved when verifying a solution is simple?*

**Context**: Often, it is hard to find a solution of a problem, but verifying whether a claimed solution is correct is simple. For example, factorizing a number is hard, but multiplying numbers is simple. Thus, when a given list of prime numbers is claimed to be the factorization of a certain number, multiplying the prime numbers and comparing the result with the certain number is a simple way of verification.

**Solution**: Solving certain problems can be speedup by first listing all possible solutions, then scanning through the list and verifying whether the current member of the list at hand is a solution or not.

The verification of the possible solutions is done via an oracle. Scanning is done by means of the Grover algorithm, thus, O(√N) invocations of the Oracle function determines the solution. A prerequisite of this pattern is that solutions can be detected by means of oracle.

**Know Uses**: Cracking keys, finding Hamiltonian cycles, solving 3-SAT, the Traveling Salesman Problem etc can be approached this way.

**Next**: The verification is performed as an *oracle*.

### 2.10. Quantum-Classic Split

**Intend**: The solution of a problem is often not achieved by only using a quantum computer. Thus, the solution is performed partially on a classical computer and partially on a quantum computer, and both parts of the solution interact.

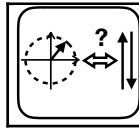 *How can a solution be split between a quantum computer and a classical computer?*

**Context**: Some quantum algorithms inherently require pre- or post-processing on a classical device, resulting in a split of the solution into a classical part and a quantum part.

Also, if a quantum computer has a low number of qubits or its gates are noisy, a solution of a problem may have to be separated into a part executed on a quantum computer and a part executed on a classical computer [22].

**Solution**: The sheer fact that a split of the algorithms may be done is important. How such a split is applied is problem dependent.

**Know Uses**: Shor's algorithm or Simon's algorithm inherently make use of classical post-processing. The algorithm in [11] to solve combinatorial optimization problems uses classical pre-processing. The algorithm of [2] uses a split into a quantum part of the solution and a classical part to enable factorization on NISQ devices.

**Next**: Data is passed from the classical part of the solution to the quantum part by proper *initialization*.

## 3. Using Patterns

### 3.1. Patterns in Software Engineering

In software engineering, pattern languages exist in a plethora of domains like object orientation, enterprise integration, cloud computing etc.. Typically, these pattern languages are delivered as books or on web pages. Software engineers determine their problem to solve and find a corresponding entry point into the pattern language. This first entry pattern links to other patterns that might be helpful in the problem context, if applicable these patterns are inspected and used too, their links are followed etc.. This way, a subgraph (aka „solution path" [27]) of the pattern language is determined that conceptually solves a complex problem. Next, the abstract solutions of the patterns of this subgraph have to be implemented (i.e. turned into concrete solutions) so that they can be executed in a computing environment.

To make this process more efficient, a pattern repository can be used [12]: in essence, a pattern repository is a specialized database that stores pattern documents and manages the links between them. It allows to query the database to find appropriate patterns (e.g. to determine the entry pattern corresponding to a problem), supports browsing the content of each pattern document, and enables navigating between patterns based on the links between them.

In practice, patterns of several domains are needed to solve a complex problem [9]. For example, building an application for a cloud environment based on microservices that must fulfill certain security requirements leans on the corresponding three pattern languages (for cloud, microservices, security). For this purpose, patterns of a pattern language of a certain domain may point to patterns of another domain.

Similarly, the pattern language for quantum algorithm can be represented in a pattern repository. The corresponding patterns may be linked with patterns from concrete quantum programming languages to support the programming of the patterns.

### 3.2. Abstract Solutions and Concrete Solutions

Patterns describe abstract solutions, independent of any concrete implementations. The advantage is that such abstract solutions fit in unforeseen contexts (new quantum hardware, new programming environments,…). But patterns represent proven solutions, i.e. by definition they are abstracted from formerly existing concrete solutions. These concrete solutions are forgotten during the act of abstraction.

By retaining (or creating) concrete solutions, making them available, and linking them with those patterns that abstract them, fosters reuse of implementations and speeds up solving problems [9]. Navigating through such enriched pattern languages allows to „harvest" concrete solutions and „glue" them together into an aggregated solution of the overall problem [10].

## 4. Outlook

We intend to grow the proposed pattern language, and make it available in our pattern repository PatternPedia [12]. In parallel, concrete implementations of the patterns in quantum languages like QASM [6] are considered, and these implementations will be linked to the corresponding patterns. Next, we want to evaluate the usefulness of the pattern language based on practical use cases.

**Acknowledgement**: I am very grateful to Johanna Barzen and Michael Falkenthal for the plethora of discussions about pattern languages and their use in different domains.